\documentclass[aps,prl,floatfix,twocolumn,reprint,amsmath,amssymb,superscriptaddress]{revtex4-1}
\usepackage{amsfonts}
\usepackage{mathrsfs}
\usepackage{amsmath}% needed for subequations
\usepackage{color}
\usepackage{natbib}
\usepackage{graphicx}
\usepackage{bm}% bold maths
\usepackage{amssymb}
\usepackage{xspace}
\usepackage{epstopdf}
\usepackage{dcolumn}% Align table columns on decimal point
\usepackage{longtable}
\usepackage{multirow}
\usepackage[colorlinks=true, letterpaper=true, pdfstartview=FitV, linkcolor=blue, citecolor=blue, urlcolor=blue]{hyperref}

\makeatletter

\newcommand{\Rmnum}[1]{\expandafter\@slowromancap\romannumeral #1@}
\makeatother
\begin{document}

\title{Mirror Protected Multiple Nodal Line Semimetals and Material Realization}

\author{Da-Shuai Ma}
\thanks{These authors contributed equally to this work.}
\affiliation{Beijing Key Laboratory of Nanophotonics and Ultrafine Optoelectronic Systems, School of Physics, Beijing Institute of Technology, Beijing 100081, China}

\author{Jianhui Zhou}
\thanks{These authors contributed equally to this work.}
\affiliation{Beijing Key Laboratory of Nanophotonics and Ultrafine Optoelectronic Systems, School of Physics, Beijing Institute of Technology, Beijing 100081, China}

\author{Botao Fu}
\affiliation{Beijing Key Laboratory of Nanophotonics and Ultrafine Optoelectronic Systems, School of Physics, Beijing Institute of Technology, Beijing 100081, China}

\author{Zhi-Ming Yu}
\affiliation{Beijing Key Laboratory of Nanophotonics and Ultrafine Optoelectronic Systems, School of Physics, Beijing Institute of Technology, Beijing 100081, China}
\affiliation{Research Laboratory for Quantum Materials, Singapore University of Technology and Design, Singapore 487372, Singapore}

\author{Cheng-Cheng Liu}
\email{ccliu@bit.edu.cn}
\affiliation{Beijing Key Laboratory of Nanophotonics and Ultrafine Optoelectronic Systems, School of Physics, Beijing Institute of Technology, Beijing 100081, China}

\author{Yugui Yao}
\email{ygyao@bit.edu.cn}
\affiliation{Beijing Key Laboratory of Nanophotonics and Ultrafine Optoelectronic Systems, School of Physics, Beijing Institute of Technology, Beijing 100081, China}

%\date{\today}
%
\begin{abstract}
The conventional  $\bm{k} \cdot \bm{p}$ method fails to capture the full and essential physics of many symmetry enriched multiple nodal line structures in the three dimensional Brillouin zone.
Here we present a new and systematical method to construct the effective lattice model of mirror symmetry protected three-dimensional multiple nodal line semimetals, when the spin-orbit interaction is ignored.
For systems with a given pair of perpendicular nodal rings, we obtain all the effective lattice models and eleven inequivalent nodal line Fermi surfaces together with their related constraints.
By means of first-principles calculations, we first propose a family of real materials, $\beta$ phase of ternary nitrides $X_{2}\mathrm{GeN_{2}}\left(X=\mathrm{Ca,Sr,Ba}\right)$, that support one kind of these novel Fermi surfaces.
Therefore, our work deepens the understanding of the nodal line structures and promotes the experimental progress of topological nodal line semimetals.
\end{abstract}
\maketitle

%%%%%%%%%%%%%%%%%%%%%%%%%%%%%%%%%%%%%%%
\textit{\textcolor{black}{Introduction.---}}Three-dimensional (3D) topological semimetals, new gapless phases of quantum matter, have attracted broad interest in  condensed matter physics and materials physics in recent years \cite{Hosur13crp,burkov2015jpcm,WengHM2016jpcm,Armitage2018RMP}.  These topological semimetals could be classified by topological invariants of the band crossing points (both type-I and -II Dirac and Weyl semimetals) or crossing lines (nodal line or ring semimetals) in the Brillouin zone (BZ). It has been shown that the nontrivial topology of discrete band crossing points of Dirac and Weyl semimetals in momentum space leads to surface Fermi arc states \cite{ WanPRB2011Weyl,Xu2015Science}, negative magnetoresistance \cite{SonSpivakPRB2013,Kim2013PRL,Huang2015PRX,Xiong2015Science,li2015NC,LiH2016NC,LiQ2016NP}, the chiral magnetic effect \cite{Fukushima2008PRD,Grushin2012PRD,zyuzin2012prb,zhou2013cpl,Vazifeh2013PRL,Goswami2013PRB,Landsteiner2014PRB,Chang2015PRBcme1,Ma2015PRB,ZhongSD2016PRLGME,OBrien2017PRL}, nonlocal transport \cite{Parameswaran2014PRX}, and other exotic electromagnetic responses \cite{Liu2013PRB,GorbarPRL2017,Liu2017PRL,Rinkel2017PRL,gooth2017nature,Nandy2017PRL,Long2018PRL}.
Recently, nodal line or ring semimetals have been predicted theoretically in a serials of materials
\cite{Burkov2011PRB,Mullen2015PRL,chen2015nanostructured,yan2016PRL,bian2016PRB,huang2016PRB,Wang2016PRL,chan2016PRB,lirg2016PRL,suny2017PRB,hirayama2017nc,Lim2017PRL,lisi2017PRB,quanY2017PRL,Ahn2017PRL,Gong2018PRL}, and some of them have been confirmed experimentally through the angle resolved photoemission spectroscopy (ARPES) in $\mathrm{PbTaSe_{2}}$ \cite{bian2016nc} and $\mathrm{ZrSiS}$ \cite{schoop2016nc,Neupane2016PRB,Wang2016aem}, through de Haas-van Alphen quantum oscillations in $\mathrm{ZrSiSe}$ and $\mathrm{ZrSiTe}$~\cite{hujin2016PRL}. Remarkably, the nearly flat drumhead-like surface states of nodal lines or rings have extremely high surface density of states and might account for the giant Friedel oscillation in the simple alkali earth metal beryllium~\cite{lirg2016PRL}.

The coexistence of different symmetries, including mirror and inversion symmetry, could give rise to multiple nodal rings and enrich the structure of Fermi surfaces of nodal line or ring semimetals~\cite{kim2015PRL,yurui2015PRL,zeng2015topological,weng2015PRB,du2016npj,bzduvsek2016nature,kobayashi2017PRB,yan2017PRB,chen2017PRB,chang2017PRB,Chang2017PRL,ezawa2017PRB,Zhong2017nc,chenW2017PRB,Mukherjee2017PRB,yu2017PRL,feng2017PRM}. Specifically, these nodal rings connect with each other in different ways, forming the nodal link \cite{Zhong2017nc,yan2017PRB,ezawa2017PRB,Chang2017PRL,chang2017PRB,chen2017PRB}, the nodal chain \cite{bzduvsek2016nature,yu2017PRL} and the nodal net \cite{feng2017PRM} that distribute in a large region of the BZ. Some simple $\bm{k} \cdot \bm{p}$ models near a single point in the BZ were constructed to describe low-energy physics of carriers near these novel Fermi surfaces \cite{Zhong2017nc,Chang2017PRL,yu2017PRL,ezawa2017PRB,feng2017PRM,zeng2015topological}. However, unlike the Dirac and Weyl semimetals,
these complicated Fermi surfaces of semimetals with multiple nodal rings or lines usually subject to symmetry constraints from multiple points in the BZ such that these simple constructions hardly capture the full and essential physics of carriers. Therefore, a systematical method to construct the effective models for the Fermi surfaces with multiple nodal rings or lines is highly desirable and is crucial in unveiling the physical properties and facilitating the material realization.

In this paper, we show a systematical procedure to construct the effective lattice models of the Fermi surfaces with mirror symmetry protected multiple nodal line structures in terms of sine and cosine functions. The effective models and corresponding inequivalent Fermi surfaces for systems with a pair of given  perpendicular nodal rings encircling a common point are obtained. We also tabulate the relevant realization conditions. In addition, we first point out that the $\beta$ phase of ternary nitrides $X_{2}\mathrm{GeN_{2}}\left(X=\mathrm{Ca,Sr,Ba}\right)$ can be candidates for one kind of such novel Fermi surfaces.

\begin{figure}[t]
\includegraphics[width=8.5cm]{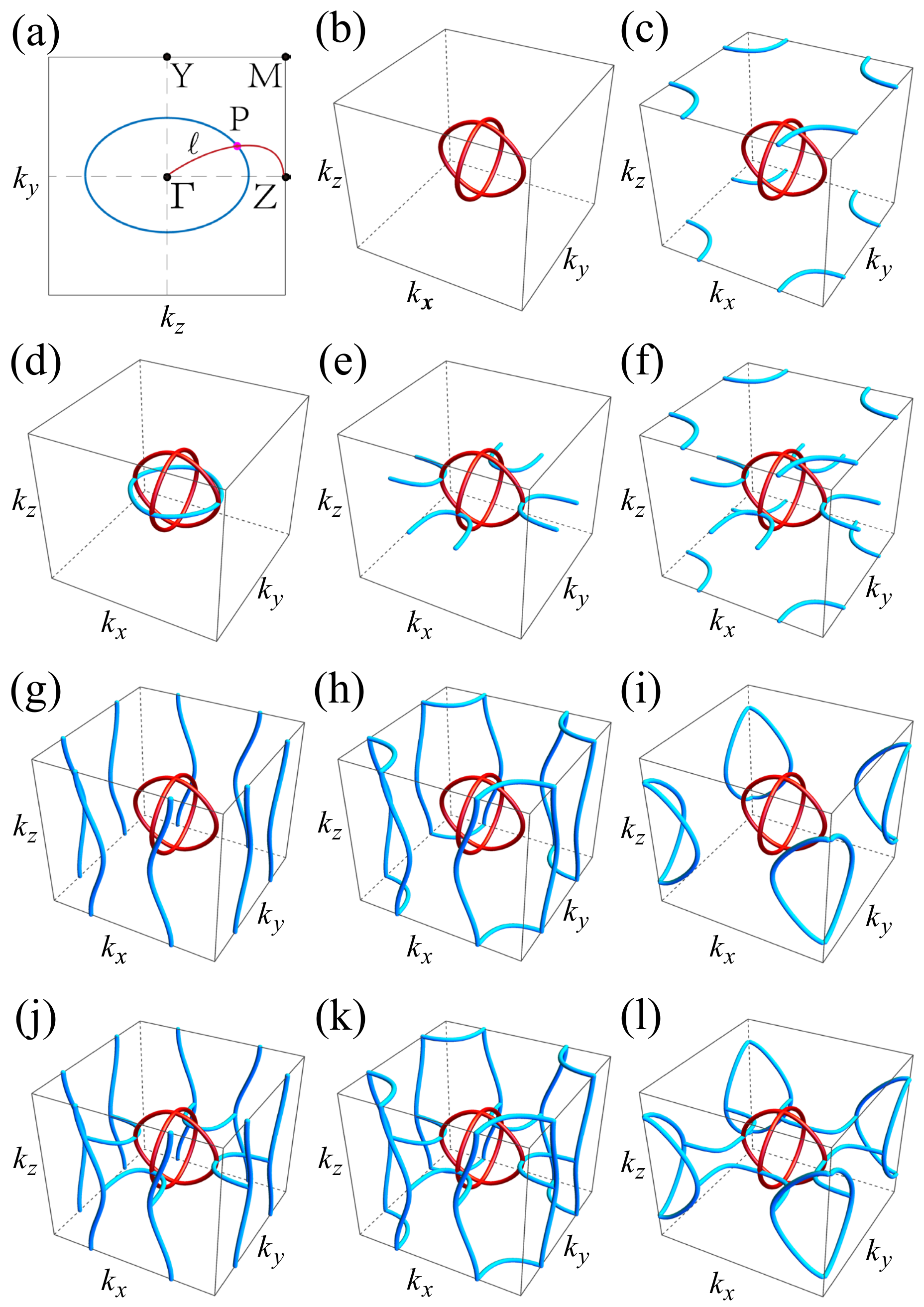}
\caption{(a) Schematic diagram of a $\mathcal{M}_{x}$  protected nodal ring in the $k_{x}=0$ plane. A band crossing point P appears on any in-plane k-path, $\ell$, that connects the $\Gamma$ point and the Z (Y, M) point.
 (b)-(l) Bird¡¯s-eye view of the possible nodal line structures obtained from Eqs. (\ref{Hami1})-(\ref{Hami3}) by setting $m=-1$  in centrosymmetric systems with mirror symmetries. Here we set $t=-8,-2$ and $0$ for the three independent regions of $t$: I, II, and III, respectively.}\label{FIG.1}
\end{figure}

\textit{\textcolor{black}{Symmetry Analysis and Effective Model.---}} Symmetries usually play a crucial role in the investigation of topological semimetals~\cite{Chiu2016RMP}.
For example, the four-fold degenerate Dirac points for Dirac semimetals require the coexistence of  inversion symmetry ($\mathcal{P}$) and time reversal symmetry ($\mathcal{T}$) \cite{Young2012PRL},  while the Weyl nodes for Weyl semimetals demand either $\mathcal{P}$ or $\mathcal{T}$ broken.
Roughly, 3D nodal line semimetals can be classified into two classes by symmetries \cite{Armitage2018RMP}. In the first class, the coexistence of $\mathcal{P}$ and $\mathcal{T}$ allows a snake like nodal line to appear in alkaline-earth compounds  $\mathrm{BaSn_{2}}$ family \cite{huang2016PRB}. The other is that a nodal line exists in the mirror invariant plane where two bands with the opposite eigenvalues of mirror or glide mirror cross with each other such as  $\mathrm{LaN}$ \cite{zeng2015topological} and  $\mathrm{PbTaSe_{2}}$ \cite{bian2016nc}.

Here, we would like to give a general method to construct the effective lattice Hamiltonian for mirror symmetry protected nodal line fermions in centrosymmetric systems with  some mirror or glide mirror symmetries. Note that SOC is ignored in the following discussions. We start from the case with two nodal rings that encircle a common point and are protected by two orthogonal mirror planes. To be specific, we suppose that these two rings, centering on the $\Gamma$ point, lie in $k_{x}=0$ and $k_{y}=0$ plane and are protected by $\mathcal{M}_{x}$ and $\mathcal{M}_{y}$, respectively (as shown in Fig. \ref{FIG.1}(b)).
The mirror operations read as
 $\mathcal{M}_{x}:(x,y,z)\rightarrow(-x+\frac{\alpha_{x}a_{x}}{2},y+\frac{\beta_{x}a_{y}}{2},z+\frac{\gamma_{x}a_{z}}{2})$, and
 $ \mathcal{M}_{y}:(x,y,z)\rightarrow(x+\frac{\alpha_{y}a_{x}}{2},-y+\frac{\beta_{y}a_{y}}{2},z+\frac{\gamma_{y}a_{z}}{2})$,
where $\alpha_{x,y}$, $\beta_{x,y}$ and $\gamma_{x,y}$ equal to $0$ or $1$, and $a_{r}$ with $r=x,y,z$ is the lattice constant along $r$-direction.
One can find that an extra symmetry operation of the combination of $\mathcal{M}_{x}$, $\mathcal{M}_{y}$ and $\mathcal{P}$, $\mathcal{P}\mathcal{M}_{y}\mathcal{M}_{x}$, is equivalent to the mirror operation about the $z$ direction $\mathcal{M}_{z}$.

\begin{table*}[t]
\caption{Parameters $\left\{\frac{m_{cz,0}}{m_{vz,0}};\xi_{x, y};\xi_{z};R\left(t\right)\right\}$ of $24$ kinds of effective lattice models and corresponding Fermi surfaces.
Because several effective Hamiltonians with different parameters may correspond to a common Fermi surface, only eleven inequivalent nodal line structures exist as shown in  Fig. \ref{FIG.1}(b)-\ref{FIG.1}(l).
}
\vspace{0.2cm}
\renewcommand\arraystretch{1.6}
\begin{tabular}{p{2.4cm}<{ \centering}p{0.9cm}<{\centering}p{0.9cm}<{\centering}p{0.9cm}<{\centering}p{0.9cm}<{\centering}p{0.9cm}<{\centering}p{0.9cm}<{\centering}p{0.9cm}<{\centering}p{0.9cm
}<{\centering}p{0.9cm}<{\centering}p{0.9cm}<{\centering}p{0.9cm}<{\centering}p{0.9cm}<{\centering}p{0.9cm}<{\centering}p{0.9cm}<{\centering}p{0.9cm
}}
\hline
\hline
 $  \frac{m_{cz,0}}{m_{vz,0}} $ & $ 1 $  & $ 1 $   & $1$ & $ 1 $  & $ 1 $   & $1$ & $ 1 $  & $ -1 $  & $ -1 $   & $-1$ & $ -1 $  & $ -1 $   & $-1$ & $ -1 $     \\
\hline
 $ \xi_{x,y} $ & $ 2 $  & $2 $   & $1,2$ & $2 $  & $1 $   & $1$ & $1$  & $1, 2 $   & $2$& $ 2 $  & $ 2 $   & $1$ & $ 1 $  & $ 1 $  \\
\hline
$ \xi_{z} $ & $1,2 $  & $1 $   & $1,2$ & $ 2 $  & $ 1 $   & $2$ & $1,2 $  & $ 1,2 $   & $1,2$& $ 2 $  & $ 1 $   & $2$ & $ 1 $  & $ 1,2 $      \\
\hline
$R(t)$ & $ \mathrm{I} $ & $\mathrm{II} $  & $\mathrm{III}$   & $ \mathrm{II} $ & $\mathrm{II} $  & $\mathrm{II}$   & $ \mathrm{I} $ & $\mathrm{III} $  & $\mathrm{I}$   & $ \mathrm{II} $ & $\mathrm{II} $  & $\mathrm{II}$ & $ \mathrm{II} $ & $\mathrm{I} $      \\
\hline
$\mathrm{Fermi~Surface}$ & $\mathrm{(b)} $ &$ \mathrm{(b)} $ &$ \mathrm{(b)} $ & $ \mathrm{(c)} $  & $\mathrm{(g)} $   &$\mathrm{(h)} $ & $\mathrm{(i)} $  & $\mathrm{(d)} $   & $\mathrm{(e)} $& $\mathrm{(e)} $ & $\mathrm{(f)} $  & $\mathrm{(j)} $   & $\mathrm{(k)} $& $\mathrm{(l)} $ \\
\hline
\hline
\end{tabular}\label{table.str}
\end{table*}

A general two-band Hamiltonian can be written as
\begin{eqnarray}
\mathcal{H}\left(\boldsymbol{k}\right)  = g_{0}\left(\boldsymbol{k}\right)\tau_{0}+\sum_{i=1}^{3} g_{i}\left(\boldsymbol{k}\right)\tau_{i},\label{Hami}
\end{eqnarray}
where $\tau_0$ is the identity matrix and $\tau_i$ with $i=1,2,3$ are the three Pauli matrices. The coefficients $ g_{i}\left(\boldsymbol{k}\right)$ with $j=0,1,2,3$ are real functions of $\boldsymbol{k}$.
To reduce the mathematical complexity, we assume that $k_{x}$ and $k_{y}$ are equivalent to each other, indicating the $ \mathcal{C}_{4z}$ symmetry.
Because $g_{0}\left(\boldsymbol{k}\right)$ does not affect the topological properties of nodal rings, we hereafter drop the $g_{0}\left(\boldsymbol{k}\right)$ term.

Now, we consider the impacts of the mirror symmetries on the Hamiltonian. Invariance of mirror symmetry $\mathcal{M}_{r}$ gives
$\mathcal{M}_{r}\mathcal{H}\left(G+q_{r}\right)\mathcal{M}_{r}^{-1}=\mathcal{H}\left(G-q_{r}\right)$,
where $q_{r}=k_{r}-G$ with $r=x,y,z$ and  $G=0$ or $\pi$. Under the basis in which both $\mathcal{H}\left(k_{r}=G\right)$ and $\mathcal{M}_{r}$ are diagonal, we can use a diagonal matrix, i.e. $\mathcal{M}_{r}=\mathrm{diag}\left[m_{cr,G},m_{vr,G}\right]$ in the mirror invariant plane, where $m_{c\left(v\right)r,G}=\pm1$ and the subscripts $c$ and $v$ denote the conduction band and valence band, respectively. After a straightforward calculation, one gets  $g_{1,2}\left(G+q_{r}\right)$ are even or odd in $q_{r}$ when $m_{cr,G}/m_{vr,G}=1$ or $-1$, while $g_{3}\left(G+q_{r}\right)$ is even in $q_{r}$ and independent of the value of $m_{cr,G}/m_{vr,G}$.

When $G=0$, a nodal ring in the $k_{r}=0$ plane requires $m_{cr,0}/m_{vr,0}=-1$, i.e. $\mathcal{M}_{r}=\tau_{3}$, with $r=x, y$. However, the matrix form of $\mathcal{M}_{z}$ still keeps undetermined. Next, we consider two separate cases: $m_{cz,0}/m_{vz,0}=1$ (\textit{Case} A) and $m_{cz,0}/m_{vz,0}=-1$ (\textit{Case} B).
For \textit{Case} A, one has $\mathcal{M}_{x,y}=\tau_{3}$ and $\mathcal{M}_{z}=\tau_{0}$, indicating that $g_{1,2}\left(\boldsymbol{k}\right)$ are odd in $k_{x,y}$ but even in $k_{z}$, and $g_{3}\left(\boldsymbol{k}\right)$ is
even in $k_{r}$.
It is known that any period function can be expanded as a linear combination of sine and cosine functions.
Thus, the periodic functions $g_{j}\left(\boldsymbol{k}\right)$ with $j=1,2,3$ can be expanded as
\begin{eqnarray}
g_{j}\left(\boldsymbol{k}\right)	&  =	& \prod_{r=x,y}\left[\sum_{n}C_{j,r,n}\mathrm{sin}\left(nk_{r}/\xi_{r}\right)\right]\nonumber\\
		&  & \times\sum_{m}C_{j,z,m} \mathrm{cos} \left(m k_{z}/\xi_{z}\right), \nonumber\\
g_{3}\left(\boldsymbol{k}\right)	 &=&	\prod_{r=x,y,z}\left[\sum_{n}C_{3,r,n} \mathrm{cos}\left(nk_{r}\right)\right],
\end{eqnarray}
where $m$ and $n$ are nonnegative integers and $\xi_{r}\equiv \frac{1}{2}\left|\frac{m_{cr,0}}{m_{vr,0}}-\frac{m_{cr,\pi}}{m_{vr,\pi}}\right|+1$ specifies the independent parities of $g_{1,2}\left(\boldsymbol{k}\right)$ nearby $k_{r}=0$ and $k_{r}=\pi$~\cite{intro}.
Similarly, for \textit{Case} B with $\mathcal{M}_{r}=\tau_{3}$, the general form of $g_{j}\left(\boldsymbol{k}\right)$ with $j=1,2$ read
\begin{eqnarray}
g_{j}\left(\boldsymbol{k}\right) & = & \prod_{r=x,y,z}\left[\sum_{n}C_{j,r,n} \mathrm{sin}\left(nk_{r}/\xi_{r}\right)\right].
\end{eqnarray}
Meanwhile, since $g_{3}\left(\boldsymbol{k}\right)$ is even in $k_{r}$ and independent of the value of $m_{cr,0}/m_{vr,0}$, $g_{3}\left(\boldsymbol{k}\right)$ in \textit{Case} B takes the same form as \textit{Case} A.
Note that $g_{1}\left(\boldsymbol{k}\right)$ does not essentially differ from $g_{2}\left(\boldsymbol{k}\right)$ except some constant coefficients.

\textit{\textcolor{black}{Nodal Line Structures.---}}
Considering a single nodal ring in the $k_{x}=0$ plane, one then has the effective Hamiltonian in this plane,
$\mathcal{H}\left(\boldsymbol{k}\right)=g_{3}\left(\boldsymbol{k}\right)\tau_{3}$.
Let us expand $g_{3}\left(\boldsymbol{k}\right)$ in terms of $\mathrm{cos}(n k_{r})$ and keep the leading terms with $n=0$ and $1$,
\begin{eqnarray}
g_{3}\left(k_{x} =0\right) & \approx & m_{0}+C_{1}^{'} \mathrm{cos}k_{y}+C_{2}^{'}\mathrm{cos} k_{z}+C_{3}^{'}\mathrm{cos}k_{y}cosk_{z}.\nonumber\
\end{eqnarray}
Since non-zero $C_{3}^{'}$ as well as different values of $C_{1}^{'}$ and $C_{2}^{'}$ only modifies the shape of the nodal ring such that one could further simplify $g_{3}\left(k_{x} =0\right)$ to $m+\mathrm{cos}k_{y}+\mathrm{cos}k_{z}$.
The corresponding energy spectrum of the Hamiltonian reads $\pm\left|g_{3}\left(k_{x} =0\right)\right|$,
which can be labeled by the eigenvalues of $\mathcal{M}_{x}=\tau_{3}$, i.e. $\pm \mathrm{sgn} \left[ g_{3}\left(k_{x} =0\right)\right]$.
In the $k_{x}=0$ plane, one has the values of $g_{3}\left(k_{x} =0\right)$ at different high symmetry points in the BZ as  $g_{3}\left(\mathrm{\Gamma}\right)=2+m$,
$g_{3}\left(\mathrm{Z}\right)=m$, $g_{3}\left(\mathrm{M}\right)=-2+m$ and $g_{3}\left(\mathrm{Y}\right)=m$.
It is clear that when $-2<m<0$, only $g_{3}\left(\mathrm{\Gamma}\right)$ is positive, the
mirror eigenvalues labeling the valence band and conduction bands switch
partners by evolving smoothly from $\Gamma$ to Z (M, Y) along any in-plane k-path $\ell$.
Consequently, a band crossing point (P) on $\ell$ always occurs.
Therefore, there exists a nodal ring that separates the $\Gamma$ point
and the other  time-reversal invariant points in the $k_{x}=0$ plane, as shown in Fig. \ref{FIG.1}(a).
Furthermore, the corresponding Hamiltonian in the $k_{y}=0$ plane can be obtained in a similar way.

Since $g_{3}\left(\boldsymbol{k}\right)$ should reduce to be $m+\mathrm{cos}k_{y,x}+\mathrm{cos}k_{z}$ in the $k_{x,y}=0$ plane, we express $g_{3}\left(\boldsymbol{k}\right)$ as follows:
{\small{}
\begin{eqnarray}
g_{3}& = & \frac{1}{t+1}\left(C+\sum_{r=x,y,z}\mathrm{cos}k_{r}+t \mathrm{cos} k_{x} \mathrm{cos}k_{y}+t\mathrm{cos}k_{z}\right),\label{Hami1}
\end{eqnarray}}%
with $C=m\left(t+1\right)-1$ and $-2<m<0$. In addition, one can expand $g_{1,2}\left(\boldsymbol{k}\right)$ and keep
the leading terms, for \textit{Case} A,
\begin{eqnarray}
g_{j}\left(\boldsymbol{k}\right)=\left[\mathrm{cos}\frac{k_{z}}{\xi_{z}}+C_{j,z}\right]\prod_{r=x,y}\mathrm{sin}\frac{k_{r}}{\xi_{r}},
\end{eqnarray}with $C_{j,z}=s_{j}\left(2-\xi_{z}\right)$ and $j=1,2$. Similarly, for \textit{Case} B one has
\begin{eqnarray}
g_{j}\left(\boldsymbol{k}\right)=\prod_{r=x,y,z}\mathrm{sin}\frac{k_{r}}{\xi_{r}}.\label{Hami3}
\end{eqnarray}

It is clear that a one-dimensional nodal line structure can exist in the $k_{r}=G$ plane only when $g_{1,2}\left(G+q_{r}\right)$ are odd functions in $q_{r}$.
Thus, one  has $\mathcal{H}\left(\boldsymbol{k}\right)=g_{3}\left(\boldsymbol{k}\right)\tau_{3}$ in all of such mirror invariant planes.
As discussed above, the nodal line structures in these planes can be well-defined by $\mathrm{sgn}\left[g_{3}\left(\boldsymbol{k}\right)\right]$.
Some straightforward calculations show that only $\mathrm{sgn}\left[g_{3}\left(\pi,\pi,0\right)\right]$ and $\mathrm{sgn}\left[g_{3}\left(\pi,\pi,\pi\right)\right]$
keep adjustable.
Specifically, one has $\mathrm{sgn}\left[g_{3}\left(\pi,\pi,0\right)\right]<0$ for $-1<t<\frac{(2-m)}{(2+m)}$ and $\mathrm{sgn}\left[g_{3}\left(\pi,\pi,\pi\right)\right]>0$ for $\frac{(4-m)}{m}<t<-1$.
It gives rise to  three independent regions of $t$:
(I) $t<\frac{(4-m)}{m}$ or $\frac{(2-m)}{(2+m)}<t$, (II) $\frac{(4-m)}{m}<t<-1$, and (III) $-1<t<\frac{(2-m)}{(2+m)}$. Summarily, the effective Hamiltonians of the possible nodal line structures can be well characterized by a set of parameters as $\left\{\frac{m_{cz,0}}{m_{vz,0}};\xi_{x,y};\xi_{z};R\left(t\right)\right\}$,
where $R\left(t\right)$ stands for the three regions of $t$ above.
Hence, one gets $24$ kinds of effective Hamiltonians as well as eleven nodal line structures in Figs. \ref{FIG.1}(b)-\ref{FIG.1}(l).
From the viewpoint of the building block, one could recognize three basic blocks among these nodal line structures: a pair of perpendicular nodal rings near the $\Gamma$ point in every figure or near the corners in the $k_z=0$ plane in Figs. \ref{FIG.1}(i) and \ref{FIG.1}(l), a single nodal ring in each mirror invariant plane perpendicular to the $k_z$ direction, and a pair of nodal line penetrating each surface BZ parallel to the $k_z$ direction as reported in \cite{chen2015nanostructured}.
We tabulate all the lattice Hamiltonians and the corresponding nodal line structures in Table \ref{table.str}.
It is one of the main results in this work.

\begin{figure}[t]
\includegraphics[width=8.5cm]{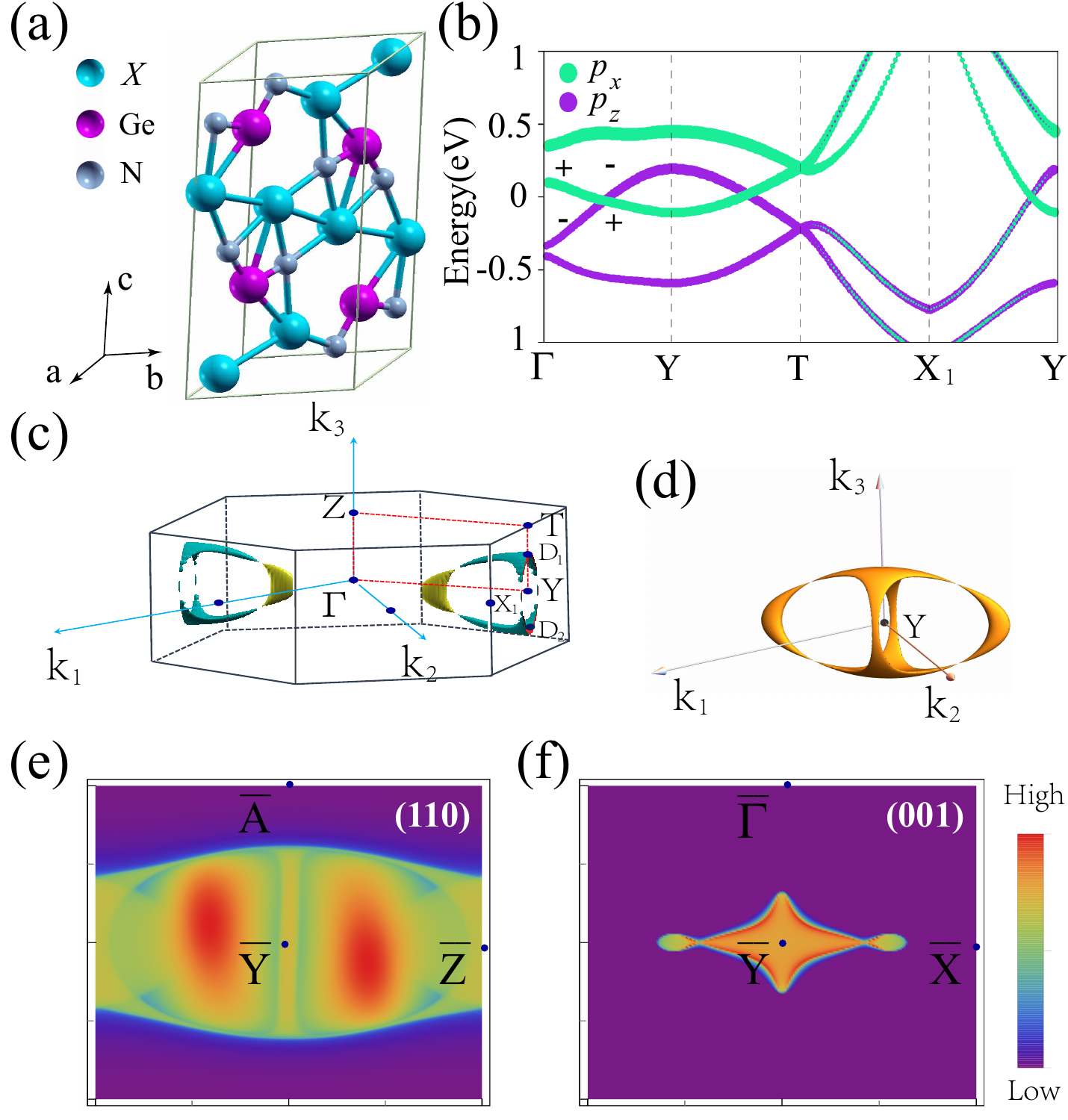}
\caption{
(a) Crystal structure of $\beta$--$X_{2}\mathrm{GeN_{2}}$.
(b) Band structure calculated by GGA-PBE of $\beta$--$\mathrm{Sr_{2}GeN_{2}}$ when SOC is ignored. The weights of the $p_{x}$ and $p_{z}$ orbits of Ge and N on valence band and conduction bands are highlighted in green and purple, respectively.
$+/-$ denote the eigenvalues of the mirror symmetry, i.e. $\mathcal{M}_{\widetilde{ab}}\colon\left(x_{1},x_{2},x_{3}\right)\longrightarrow\left(-x_{2},-x_{1},x_{3}\right)$, where $x_{i}$ is the position under the basis $\left<a,b,c\right>$.
(c) The Fermi surface of $\beta$--$\mathrm{Sr_{2}GeN_{2}}$ indicates two nodal lines perpendicular to each other near the Y point in the first BZ.
(d) The Fermi surface obtained from our model Hamiltonian with parameters $\left\{1;1;1;\mathrm{III}\right\}$~\cite{SMsoc}.
(e) Isoenergy band contours of $\left(110\right)$ surfaces at $20$ $\mathrm{meV}$ relative to the Fermi level.
A surface flat band inside the projected circle indicates the existence of a drumhead-like surface state.
(f) Isoenergy band contours of $\left(001\right)$ surface. Only a cross formed by the projected bulk states remains in the surface BZ.
}\label{FIG.3}
\end{figure}

\textit{\textcolor{black}{Material Realization.---}}Let us turn to discuss the concrete material realization of these nontrivial Fermi surfaces.
In fact, the nodal line structure in Fig. \ref{FIG.1}(d), has been predicted in $\mathrm{Cu_{3}PdN}$ \cite{kim2015PRL,yurui2015PRL}, rare earth monopnictide $\mathrm{LaN}$ \cite{zeng2015topological}, 3D graphene networks \cite{weng2015PRB}, and $\mathrm{CaTe}$ \cite{du2016npj}.
In this work, we first demonstrate that $\beta$ phase of ternary nitrides $X_{2}\mathrm{GeN_{2}}\left(X=\mathrm{Ca,Sr,Ba}\right)$ whose crystal structures are depicted in Fig. \ref{FIG.3}(a) possess the novel nodal line structure in Fig. \ref{FIG.1}(b).
$\beta$--$\mathrm{Sr_{2}GeN_{2}}$ had been synthesized experimentally in 2005~\cite{park2005synthesis}.
And our first-principles calculations of phonon spectra show that $\beta$--$\mathrm{Ca_{2}GeN_{2}}$ and $\beta$--$\mathrm{Ba_{2}GeN_{2}}$  are dynamically stable~\cite{SMsoc}.
We shall take $\beta$--$\mathrm{Sr_{2}GeN_{2}}$ as an example to investigate the nontrivial nodal line structure in $\beta$--$X_{2}\mathrm{GeN_{2}}$.

The energy band structure of $\beta$--$\mathrm{Sr_{2}GeN_{2}}$ without the SOC effect is depicted in Fig. \ref{FIG.3}(b)~\cite{SMsoc}.
The  band-project analysis reveals that both the valence band and conduction band are mainly composed by $p_{x}$ and $p_{z}$ orbits of Ge and N atoms, indicating a tiny SOC effect.
Further band structure calculations demonstrate that there are two nodal rings in $\mathrm{\Gamma ZTY}$  and $\mathrm{TYX_{1}A_{1}}$ plane, respectively.
The Fermi surface in Fig. \ref{FIG.3}(c) gives a visual recognition of the unique nodal line structure. Both of the two nodal rings center on the Y point and connect with each other at two points $\mathrm{D}_{1}$  and $\mathrm{D}_{2}$ that locate above and below the Y point along the $k_{3}$ axis.
Fig. \ref{FIG.3}(d) clearly shows that the effective Hamiltonian with parameters $\left\{1;1;1;\mathrm{III}\right\}$ could produce the Fermi surface in Fig.~\ref{FIG.3}(c).

The remarkable feature of a topological nodal ring is the drumhead-like state in the surface BZ,
which can be probed by various experimental techniques, such as the scanning tunneling microscope and the ARPES.
For the $\left(110\right)$ surface in Fig. \ref{FIG.3}(e), a surface flat band resides inside the projected circle. The other nodal ring will be a line in Fig. \ref{FIG.3}(e) as the two nodal rings are perpendicular to each other. Nevertheless,  since both of the two nodal rings perpendicular to $\left(001\right)$ surface, the isoenergy band contour of the $\left(001\right)$ surface will be a cross just as shown in Fig. \ref{FIG.3}(f). Thus, the detection of the surface states in this system would be strongly affected by its crystal orientation.

\textit{\textcolor{black}{Conclusion.---}}In summary, we introduced a systematical method to build the effective Hamiltonians of 3D multiple nodal line semimetals. It has been shown that
there exist eleven distinct kinds of symmetry protected nodal line structures in systems with inversion symmetry and perpendicular crystalline mirror planes when SOC is ignored.
These corresponding effective Hamiltonians can be fully determined by four parameters.
In addition, we first found that $\beta$--$X_{2}\mathrm{GeN_{2}}\left(X=\mathrm{Ca}\mathrm{,Sr,Ba}\right)$ are candidates for 3D semimetals with the novel nodal line structure in Fig. \ref{FIG.1}(b).

It should be emphasized that our method can be applicable to bosonic systems as well, such as photons and phonons. The recent experimental advances of nodal line and nodal chain semimetal in photonic crystals \cite{gao2018nc,yan2018np} would inspire the investigation of these new nodal line structures in bosonic systems. Furthermore, our idea can be generalized to systems with and without $\mathcal{T}$.

This work was supported by the National Key R\&D Program of China (No. 2016YFA0300600), the National Natural Science Foundation of China (Nos. 11774028, 11734003, 11574029, 11404022),
the MOST Project of China (No. 2014CB920903),  and Basic Research Funds of Beijing Institute of Technology (No. 2017CX01018).

\onecolumngrid

\section*{Supplementary material for ``Mirror Protected Multiple Nodal Line Semimetals and Material Realization"}

This supplemental material contains the first-principles calculations of the band structures of $\beta$--$X_{2}\mathrm{GeN_{2}}\left(X=\mathrm{Ca,Sr,Ba}\right)$, the impact of the SOC effect, constructions of the effective Hamiltonians, and the dynamical stability.

\textit{\textcolor{black}{Methods and Material.---}}
To clearly understand the electronic properties of $\beta$--$X_{2}\mathrm{GeN_{2}}$, we performed first-principles calculations by using the Vienna $ab$-$initio$ simulation package (VASP) based on the generalized gradient approximation (GGA) \cite{Perdew1996PRL} in the Perdew-BurkeErnzerhof (PBE)  exchange-correlation functional \cite{kresse1994PRB,kresse1996PRB}. A $\Gamma$-centered k-mesh $7\times7\times5$ is used to sample the Brillouin zone.
The nonlocal HeydScuseria-Ernzerhof hybrid functional calculations (HSE06) \cite{heyd2003jcp,heyd2006jcp} are also used to check the band structures.
The maximally localized Wannier functions (MLWF) \cite{mostofi2008cpc,marzari2012rmp} projected from the bulk Bloch wave functions and wann\_tools \cite{wu2017cpc} are used to extract the topological surface states.

\begin{figure}[h]
\includegraphics[width=16cm]{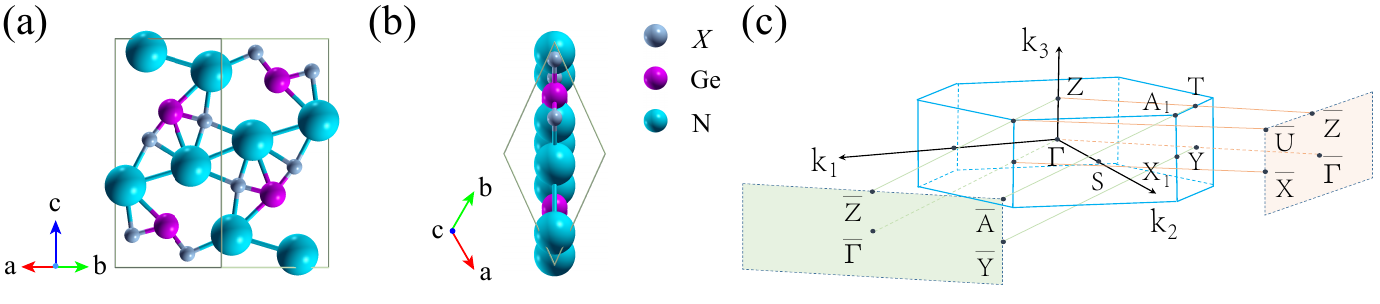}
\caption{
(a) Side-view and (b) Top-view of the crystal structure of $\beta$--$X_{2}\mathrm{GeN_{2}}\left(X=\mathrm{Ca,Sr,Ba}\right)$.
All the atoms can be found in the $(110)$ plane. Here, the balls in sapphire, purple and gray represent $X$, Ge and N atoms, respectively.
(c) Bulk Brillouin zone is given in sapphire, and the projected surface Brillouin zones of $(110)$ and $(\overline{1}10)$ planes are given in shadow.
}\label{S1}
\end{figure}

The crystal structure of $\beta$--$X_{2}\mathrm{GeN_{2}}$  is shown in Fig. \ref{S1}(a)-Fig. \ref{S1}(b).
This crystal structure belongs to the Cmce space group (No. $64$).
The lattice parameters $a=b=6.369\mathring{\mathrm{A}}$ and $c=12.312\mathring{\mathrm{A}}$ are fully optimized.
The angle between lattice vectors $a$ and $b$ is about $129.3^{\circ}$.
This crystal structure possesses three independent symmetries, i.e.
one mirror symmetry $\mathcal{M}_{\widetilde{ab}}\colon\left(x_{1},x_{2},x_{3}\right)\longrightarrow\left(-x_{2},-x_{1},x_{3}\right)$ and two glide mirror symmetries $\mathcal{M}_{ab}\colon\left(x_{1},x_{2},x_{3}\right)\longrightarrow\left(x_{2}+1/2,x_{1}+1/2,x_{3}+1/2\right)$ and $\mathcal{M}_{c}\colon\left(x_{1},x_{2},x_{3}\right)\longrightarrow\left(x_{1}+1/2,x_{2}+1/2,-x_{3}+1/2\right)$, where $x_{i}$ is the position under the basis $\left<a,b,c\right>$.
Combining these three symmetries in different means, one could have identity symmetry, inversion symmetry, one twofold rotation symmetry, and two twofold screw symmetries.
In addition, the corresponding bulk and surface Brillouin zones are depicted in Fig. \ref{S1}(c).

\textcolor{black}{\textit{Energy Band Structures.}---} As discussed in the main text, the energy band structure  of $\beta$--$X_{2}\mathrm{GeN_{2}}$ obtained by GGA-PBE indicates this family material can be candidates for multiple nodal line semimetals consisting of  two perpendicular nodal rings encircling the Y point.
However, the GGA-PBE calculation sometimes can not produce the precise band gap.
%In other words, the real system may be a insulator even when the GGA-PBE calculation indicates semimetal phase.
%
\begin{figure}[h]
\includegraphics[width=17cm]{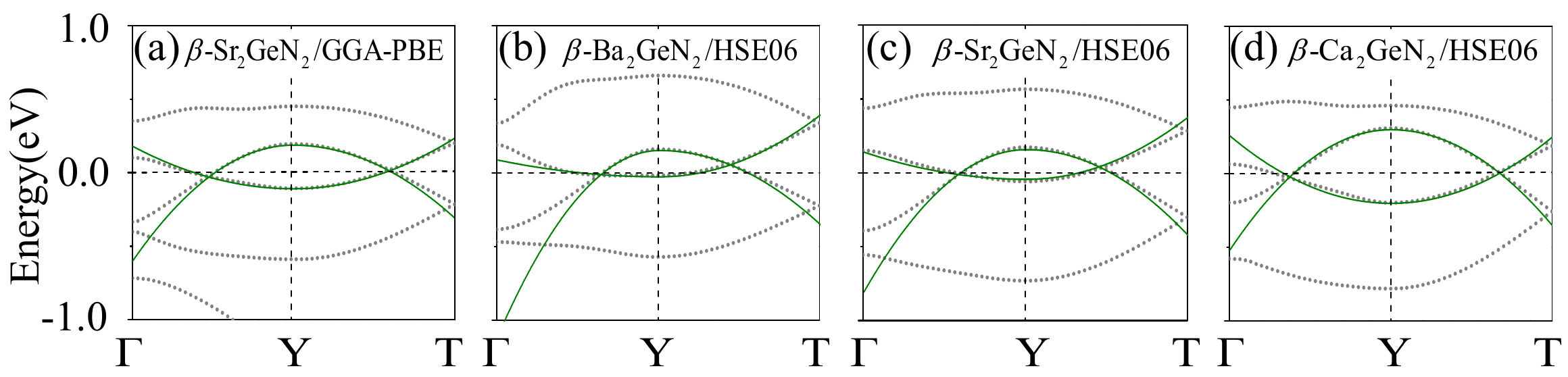}
\caption{
(a) Band structure of  $\beta$--$\mathrm{Sr_{2}GeN_{2}}$ calculated by GGA-PBE.
(b)-(d) Band structure calculated by HSE06 of  $\beta$--$\mathrm{Ba_{2}GeN_{2}}$ and $\beta$--$\mathrm{Sr_{2}GeN_{2}}$ with a 2\% compress strain and $\beta$--$\mathrm{Ca_{2}GeN_{2}}$.
All of the $\beta$--$X_{2}\mathrm{GeN_{2}}\left(X=\mathrm{Ca}\mathrm{,Sr,Ba}\right)$
( 2\% compress strain for $\beta$--$\mathrm{Ba}_{2}\mathrm{GeN_{2}}$ and $\beta$--$\mathrm{Sr}_{2}\mathrm{GeN_{2}}$) have the similar band structures as $\beta$--$\mathrm{Ba_{2}GeN_{2}}$ calculated by GGA-PBE.
The  band structures from  $\bm{k} \cdot \bm{p}$ Hamiltonian Eqs. (\ref{Hami})-(\ref{Hami1}) with parameters in Table \ref{table.1} are plotted in green.}\label{S2}
\end{figure}
Hence, we perform HSE06 calculations to check the band structures of $\beta$--$X_{2}\mathrm{GeN_{2}}$.
The HSE06 results in Fig. \ref{S2}(b)-\ref{S2}(d) demonstrate that $\beta$--$\mathrm{Ca_{2}GeN_{2}}$ supports the multiple nodal line structure. The existences of multiple nodal line structures in $\beta$--$\mathrm{Ba_{2}GeN_{2}}$ and $\beta$--$\mathrm{Sr_{2}GeN_{2}}$ require a slight compress strain, e.g. 2\%.
%Fortunately, a slight compress strain (2\%) can drives the novel line-nodal structure to $\beta$--$\mathrm{Ba_{2}GeN_{2}}$ and $\beta$--$\mathrm{Sr_{2}GeN_{2}}$.
%

\textit{\textcolor{black}{SOC Effect.---}} We turn to discuss impacts of the SOC effect on the energy band structures of $\beta$--$X_{2}\mathrm{GeN_{2}}$.
Since a mirror operation acts both on the spatial and the spin spaces, a mirror operation, e.g.  $\mathcal{M}_{z}$  can flips  the $x$  and $y$  components of spin but the spin component along the mirror axis remains intact.
Hence, the eigenvalue of mirror symmetry $\mathcal{M}_{z}$ reads $m_{z}^{2}=\left(-1\right)^{2S}$, where $S=1/2$ and $1$ refer to the cases with and without SOC effect, respectively.
The doubly degenerate conduction or valence band should be labeled by $m_{z}(i , -i)$.
Since the energy bands with the same eigenvalues of mirror symmetry would repulse each other.
As a result, the mirror symmetry protected nodal line when the SOC effect is neglected will be gapped out when the SOC effect is taken into account.

\begin{figure}[h]
\includegraphics[width=10cm]{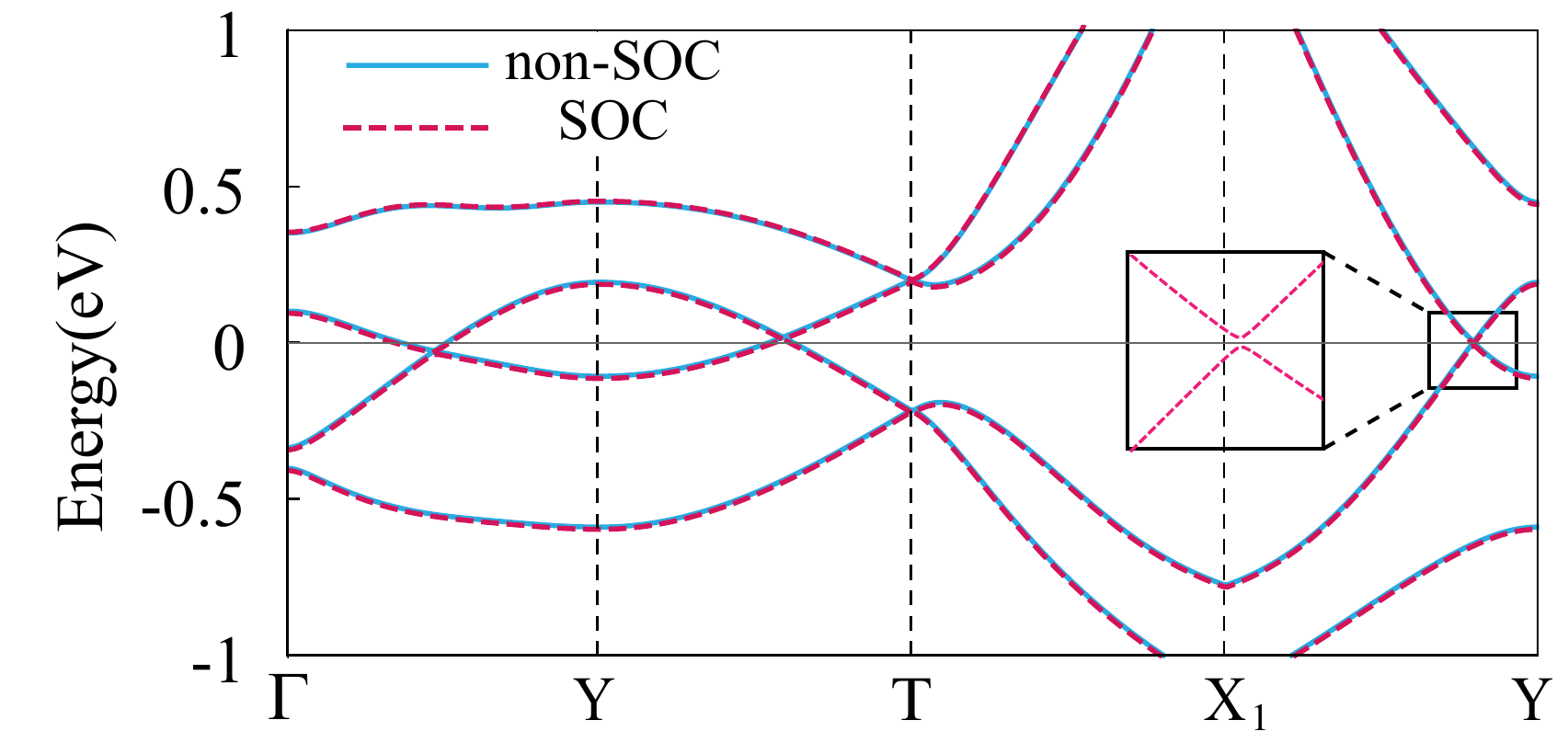}
\caption{
Comparison of Band structures of $\beta$--$\mathrm{Sr_{2}GeN_{2}}$ with and without SOC effect. The negligible SOC effect only leads to a small correction to the band structures of  $\beta$--$\mathrm{Sr_{2}GeN_{2}}$.}\label{ss2}
\end{figure}

Since the relevant bands mainly consist of $p_{x}$ and $p_{z}$ orbital of the light elements N and Ge, the SOC effect in $\beta$--$X_{2}\mathrm{GeN_{2}}$ may be negligible.
We show the band structure of  $\beta$--$\mathrm{Sr_{2}GeN_{2}}$ with and without SOC effect in Fig. \ref{ss2}. The results  indicates that the band crossing along the two nodal rings gapped out slightly (less than $10~\mathrm{meV}$). Hence, the family of materials even with the SOC effect can be regarded as nodal line semimetals as proposed in the main text.

\begin{table*}[h]
\caption{Parameters of the $\bm{k} \cdot \bm{p}$ Hamiltonian fitted by the first-principles calculations. The percentages in this Table indicate the strength of compress strain.}
\vspace{0.2cm}
\renewcommand\arraystretch{1.6}
\begin{tabular}{p{3.8cm}<{\centering}p{1.4cm}<{\centering}p{1.4cm}<{\centering}p{1.4cm}<{\centering}p{1.4cm}<{\centering}p{1.4cm}<{\centering}p{1.4cm}<{\centering}p{1.4cm}<{\centering}p{1.4cm}<{\centering}p{1.4cm}<{\centering}}
\hline
\hline
   & $a_{0}$  & $a_{1}$  & $a_{2}$  & $a_{3}$  & $ b $  & $d_{0}$  & $d_{1}$  & $d_{2}$  & $d_{3}$  \\
     System/ Method/ Strain& $(\mathrm{eV})$  & ($\mathrm{eV\cdot\mathring{A}^{2}})$  & ($\mathrm{eV\cdot\mathring{A}^{2}})$  & $(\mathrm{eV\cdot\mathring{A}^{2}})$  & $(\mathrm{eV\cdot\mathring{A}^{2}})$  & $(\mathrm{eV})$  & $(\mathrm{eV\cdot\mathring{A}^{2}})$  & $(\mathrm{eV\cdot\mathring{A}^{2}})$  & $(\mathrm{eV\cdot\mathring{A}^{2}})$  \\
\hline
 $\beta$--$\mathrm{Sr_{2}GeN_{2}}$/ $\mathrm{GGA}$/ $0\%$  & 0.045  & -6.28  & -0.25  & -0.28  & -4.25 & -0.15  & 0.54  & 23.57  & 1.67   \\
 \hline
 $\beta$--$\mathrm{Ba_{2}GeN_{2}}$/ $\mathrm{HSE06}$/$2\%$ & 0.069  & 38.94  & -0.58  & -0.17  & -8.16 & -0.09  & 0.69  & 134.4  & 1.87   \\
 \hline
 $\beta$--$\mathrm{Sr_{2}GeN_{2}}$/ $\mathrm{HSE06}$/ $2\%$  & 0.063  & -11.39  & -0.39  & -0.34  & -4.74 & -0.10  & 0.57  & 35.78  & 2.03   \\
 \hline
 $\beta$--$\mathrm{Ca_{2}GeN_{2}}$/ $\mathrm{HSE06}$/ $0\%$ & 0.052  & -9.74  & -0.18  & -0.42  & -5.88 & -0.25  & 0.64  & 49.08  & 2.21   \\
\hline
\hline
\end{tabular}\label{table.1}
\end{table*}

\textit{\textcolor{black}{Model Hamiltonian.---}}
It has been shown that $\beta$--$X_{2}\mathrm{GeN_{2}}$  possesses the nodal line structure schematically shown in Fig. 1(b) in the main text.
Based on the eigenvalues of the mirror symmetries from the first-principles calculations, we only need to expand lattice model Hamiltonian
\begin{eqnarray}
\mathcal{H}\left(\boldsymbol{k}\right)  = g_{0}\left(\boldsymbol{k}\right)\tau_{0}+g_{1}\left(\boldsymbol{k}\right)\tau_{1}+g_{2}\left(\boldsymbol{k}\right)\tau_{2}+g_{3}\left(\boldsymbol{k}\right)\tau_{3},\label{Hami}
\end{eqnarray}
and remain the second order terms in $k_{\alpha}$ with $\alpha=x,y,z$. In the case with parameters $\left\{ 1;1;1;\mathrm{III}\right\}$, the corresponding coefficients are given as
\begin{eqnarray}
g_{1}\left(\boldsymbol{k}\right) =g_{2}\left(\boldsymbol{k}\right)=  \mathrm{sin}k_{x} \mathrm{sin}k_{y},	& \ \ \
&g_{3} =  \mathrm{cos}k_{x}+\mathrm{cos}k_{y}+\mathrm{cos}k_{z}+m.
\end{eqnarray}
Moreover, the time-reversal symmetry takes the form as $\widehat{T}=\mathcal{K}\tau_{0}$, and $\mathcal{K}$ is the complex conjugate operator.
$\widehat{O}\mathcal{H}\left(\boldsymbol{k}\right)\widehat{O}^{-1}=\mathcal{H}\left(-\boldsymbol{k}\right)$ leads to $g_{2}=0$,
where the $\widehat{O}$ is the inversion operation or time-reversal operation.
Then, the remaining coefficients of the effective $\bm{k}\cdot\bm{p}$ Hamiltonian are
 \begin{eqnarray}
g_{0}=a_{0}+a_{1}k_{x}^{2}+a_{2}k_{y}^{2}+a_{3}k_{z}^{2},\ \ \ g_{1}=bk_{x}k_{y},\ \ \ g_{3}=d_{0}+d_{1}k_{x}^{2}+d_{2}k_{y}^{2}+d_{3}k_{z}^{2},\label{Hami1}
\end{eqnarray}
with $k_{x}=k_{1}+k_{2}$ , $k_{y}=k_{2}-k_{1}$ and $k_{z}=k_{3}$.
Here these parameters $a_{j}$, $b$ and $d_{j}$ with $j=0,1,2,3$ are used to fit the band structures of real materials and given in Table \ref{table.1}.
%, $g_{0}$ is recalled for the weak energy dispersion along the nodal line structure.
%Meanwhile, the extra parameters are introduced to fit the real system, and the parameters of the model Hamiltonian are given by Table \ref{table.1}.
The corresponding Fermi surface from this $\bm{k}\cdot\bm{p}$ Hamiltonian is shown in Fig. 2(d) in the main text.

In the $k_{x}=0$ plane and $k_{y}=0$ plane, we can get two functions, i.e. $d_{2}k_{y}^{2}+d_{3}k_{z}^{2}=-d_{0}$ and $d_{1}k_{x}^{2}+d_{3}k_{z}^{2}=-d_{0}$
from the equality of the two eigenvalues of the Hamiltonian.
When $d_{j}d_{0}<0$ with $j=1,2,3$, each of the two functions can determine an ellipse, giving rise to two nodal rings in the $\mathcal{M}_{y}$
and $\mathcal{M}_{x}$ planes.
Moreover, the two ellipses cross with the $k_{3}$ axis at $\mathrm{D}_{1,2}=\left(0,0,\pm\sqrt{-d_{0}/d_{3}}\right)$ points.
In other words, the two nodal rings connect each other at $\mathrm{D}_{1,2}$ located above and below the Y point in Fig. 2(c) in the main text.

\begin{figure}[h]
\includegraphics[width=15cm]{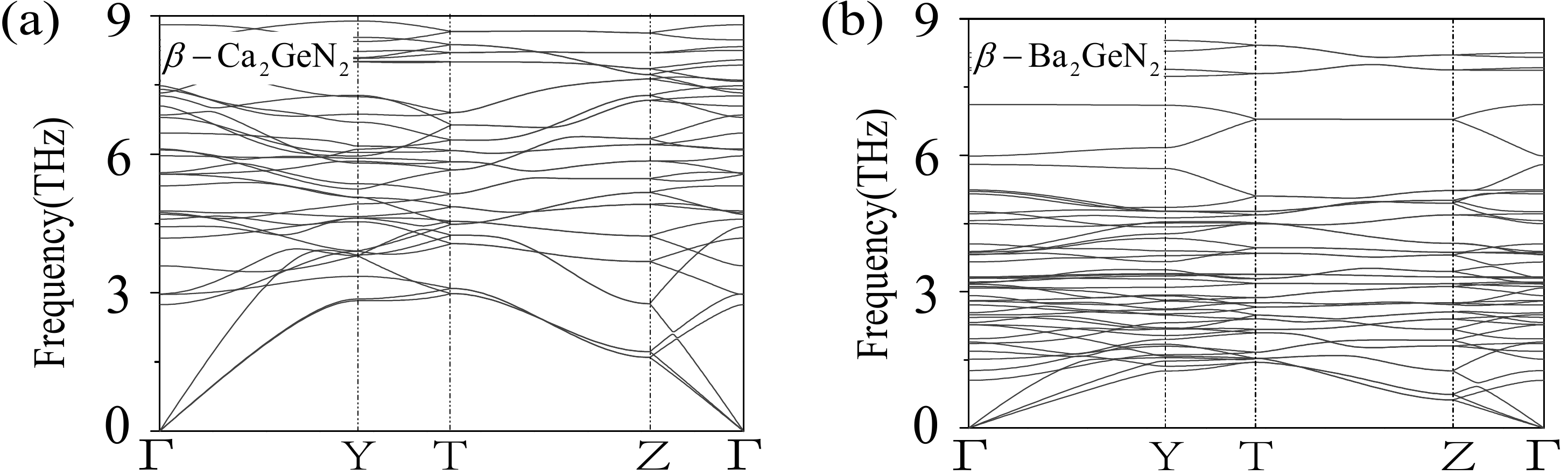}
\caption{Phonon spectra of $\beta$--$\mathrm{Ba_{2}GeN_{2}}$ in (a) and $\beta$--$\mathrm{Ca_{2}GeN_{2}}$ in (b) indicate their stabilities.}\label{FIG.S1}
\end{figure}

\textit{\textcolor{black}{Dynamical Stability.---}} In fact, $\beta$--$\mathrm{Sr_{2}GeN_{2}}$ had been successfully synthesized in experiments \cite{park2005synthesis}.
So far, there is no experimental synthesis of $\beta$--$\mathrm{Ca_{2}GeN_{2}}$ and $\beta$--$\mathrm{Ba_{2}GeN_{2}}$.
Here, we would like to study the stability of $\beta$--$\mathrm{Ca_{2}GeN_{2}}$ and $\beta$--$\mathrm{Ba_{2}GeN_{2}}$ by calculating their phonon spectra.
As shown in Fig. \ref{FIG.S1}, no imaginary frequency point in the phonon spectra suggests that both two materials are dynamically stable and may be synthesized in laboratories.

\twocolumngrid


\begin{thebibliography}{72}
\bibitem{intro} The parities of $g_{1,2}\left(\boldsymbol{k}\right)$ nearby $k_{r}=0$ and $k_{r}=\pi$ are defined by the eigenvalues of mirror symmetries and independent.
One thus needs to extend the periods of $g_{1,2}\left(\boldsymbol{k}_{r}\right)$ from $2\pi/a_{r}$ to $4\pi/a_{r}$ when $g_{1,2}\left(\boldsymbol{k}\right)$ have opposite parities nearby $k_{r}=0$ and $k_{r}=\pi$. These two distinct periods can be unified by introducing a parameter $\xi_{r}=\frac{1}{2}\left|\frac{m_{cr,0}}{m_{vr,0}}-\frac{m_{cr,\pi}}{m_{vr,\pi}}\right|+1$, as $2\xi_{r}\pi/a_{r}$. Physically, for $\xi_{r}=1$, both the $k_{r}=0$ and $k_{r}=\pi$ planes possess the same eigenvalues of the mirror operation $\mathcal{M}_{r}$. In other words, each of the two planes would support a nodal ring, otherwise no mirror protected nodal ring exists in both planes. For $\xi_{r}=2$, at most one of the two planes supports a mirror protected nodal line.
Noted that this extension of period makes the energy dispersion satisfy the fundamental relation $E_{n}\left(\boldsymbol{k}\right)=E_{n}\left(\boldsymbol{k}+\boldsymbol{K}_{r}\right)$.
\end{thebibliography}
\end{document}